\documentclass[prl,aps,twocolumn,preprintnumbers,superscriptaddress]{revtex4-2}
\usepackage{amsmath}
\usepackage{color}
\usepackage{graphicx}
\usepackage{hyperref}
\hypersetup{colorlinks, citecolor=blue, linkcolor=black, urlcolor=blue}
\usepackage[caption=false]{subfig}

\usepackage[american]{babel}
\usepackage{microtype}
 \usepackage[table,xcdraw]{xcolor}
\definecolor{darkblue}{rgb}{0.0,0.0,0.6}

\usepackage{geometry}
\usepackage{fullpage}
\begin{document}

\title{ Primordial Black-Hole Dark Matter via Warm Natural Inflation}

\author{Miguel~Correa}
\email{mcorrea2@nd.edu}
\affiliation{Center for Astrophysics,
Department of Physics and Astronomy, University of Notre Dame, Notre Dame, IN 46556, USA.}
\author{Mayukh R. Gangopadhyay}
\email{mayukhraj@gmail.com}
\affiliation{Centre For Cosmology and Science Popularization (CCSP), SGT University, Gurugram, Delhi- NCR, Haryana- 122505, India.}
\author{Nur Jaman}
\email{nurjaman.pdf@iiserkol.ac.in }
\affiliation{Department of Physical Sciences, \\
 Indian Institute of Science Education and Research Kolkata, Mohanpur - 741 246, WB, India}
\author{Grant J. Mathews}
\email{gmathews@nd.edu}
\affiliation{Center for Astrophysics,
Department of Physics and Astronomy, University of Notre Dame, Notre Dame, IN 46556, USA.}

\begin{abstract}
We report on a study of the natural warm inflationary paradigm (WNI). We show two important new results arise in this model.  One is that  the observational constraints on the primordial power spectrum from the cosmic microwave background (CMB) can be satisfied without going beyond the Planck scale of the effective field theory. The second is that WNI can inevitably provide perfect conditions for the production of primordial black holes (PBHs) in the golden window of black-hole mass range ($10^{-16}- 10^{-11} M_{\odot}$) where it can account for all of the the dark matter content of the universe while satisfying observational constraints.
\end{abstract}
\maketitle
\section{Introduction}
In addition to addressing  inherent inconsistencies in the standard cosmological model such as the flatness and horizon problems, inflation remains as one of the most promising means to generate Gaussian and adiabatic perturbations in the early universe \cite{inflation,lindinflation,inflation2,starobinsky}.  
In the standard cold inflationary paradigm (CI), the inflaton is a self-coupled ``slow-rolling"  scalar field with a nearly flat effective potential that generates a (quasi) exponential cosmic expansion. In this case, the temperature becomes negligibly small until the epoch of reheating. In another dynamical realization, called warm inflation (WI), the inflaton has non-negligible interactions with preexisting matter fields such as radiation. This leads to a rich phenomenology in the case of WI \cite{wi1,wii1, wi2,wi9, wi3, wi4, wi5, wi6, wi7, wi8}.

Any inflationary paradigm is characterized by the effective potential of the inflaton field.  In this paper we consider an inflaton effective potential produced by symmetry breaking as  proposed in Ref.~\cite{Freese}. In this ``natural inflation" model, the field is essentially a Goldstone boson with an approximate shift symmetry. Although the flatness of the potential is ensured via this symmetry, current cosmological observations~\cite{pl1, pl2, wmap, keck, bkplanck} do not favour the ``cold" natural inflation scenario.
In this letter, however, we  have studied the dynamics of natural inflation in the WI paradigm.  One important new result  demonstrated here is that warm natural inflation is consistent with the observational constraints from the CMB for a symmetry breaking scale that falls  below the effective Planck scale, thereby maintaining the original theoretical motivation for natural inflation.

A second important aspect of the present work concerns primordial black holes (PBHs) which may have formed in the  early universe~\cite{Zeldovich:1967lct,Hawking:1971ei,Carr:1974nx}, and are of great interest from both a theoretical and observational perspective. 
%PBHs represent a unique probe for the  study of the small-scale of the early universe since they would have formed from large perturbation generated during inflation. The study of PBHs has also gained considerable importance in the light of the recent observations \cite{ligo1, ligo2, ligo3, ligo4} of gravitational waves (GWs). The most intriguing  possibility, which we demonstrate here, is that the massive PBHs could account for all or a fraction of the dark matter (DM) content of the universe if their masses fall in the  golden window of ($10^{-16}- 10^{-11} M_{\odot}$)~\cite{Chapline:1975ojl,Carr:2016drx,Carr:2020xqk,Ballesteros:2017fsr}. 
PBHs represent a unique probe for the  study of the small-scale of the early universe since they would have formed from large perturbations generated during inflation. The study of PBHs has also gained considerable importance in the light of the recent observations \cite{ligo1, ligo2, ligo3, ligo4} of gravitational waves (GWs). The most intriguing  possibility, which we demonstrate here, is that the massive PBHs could account for all or a fraction of the dark matter (DM) content of the universe if their masses fall in the suitable window allowed by observations ~\cite{Chapline:1975ojl,Carr:2016drx,Carr:2020xqk,Ballesteros:2017fsr,Belotsky:2014kca,Belotsky:2018wph,Carr:2020gox}. In the so called golden window ($M_{\rm PBH}\sim $ $10^{-16}- 10^{-11} M_{\odot}$) it is possible to have all DM from PBHs.
Many attempts have been made in the context of CI to form PBHs.  However,  in most cases one needs to introduce some kind of inflection point in the potential during the slow-roll. This provides the increment in fluctuations which is required for  PBH production when the relevant mode reenters the horizon at later stages. Although the idea of such a production mechanism is intriguing,  it requires much fine-tuning. In Ref.~\cite{arya}, however,  it was shown for the first time that in the case of WI, PBH production is quite natural.  This is because the enhancement of the power spectrum is inevitable in WI dynamics. This claim was solidified in \cite{marpbh}.  However, in both of the previous works the predicted mass range of PBHs was not within allowed limits set by PBH lifetimes and observations. Rather, in both cases, the produced PBHs have quite a low mass and  would have  evaporated in the early Universe. 

In this letter, however, we show for the first time, the novel result that the desired mass range of PBHs can be achieved in WI by using the theoretically well motivated natural inflation effective potential. In this case,  the inherent shift symmetry of the potential along with the warm dynamics leads  to the production of  PBHs ``naturally" and could account for the entire dark matter content of the universe.
%%%%%%%%%%%%%%%%%%%%%%%%%%%%%%%%%%%%%%%%%%%%%%%%%%%%%%%%%%%%%%%%%%%%%%%%%%%%
\section{Warm Natural Inflation (WNI) Dynamics:}
The physical dynamics during the inflationary epoch in the WI scenario, is basically characterized by three background equations describing the evolution of the inflaton field, the evolution of matter, and the cosmic expansion.  These are:
\begin{equation}
\ddot{\phi}(t)+3H(1+Q)\dot{\phi}(t)+ V_{, \phi}=0~~,
\label{eqn1}
\end{equation}
\begin{equation}
\dot{\rho}_R+4H\rho_R= 3HQ\dot{\phi}^2~~,
\label{eqn2}
\end{equation}
\begin{equation}
H^2 = \frac{1}{3M_P^2}(\rho_{\phi}+\rho_R)~~, 
\label{eqn3}
\end{equation}
where, $V_{,\phi}\equiv \frac{\partial V}{\partial \phi}$ and overdots represents reperent derivatives w.r.t cosmic time.
Here, the only new parameter w.r.t.~the CI scenario is the dissipation ratio $Q$ where $Q\equiv {\Upsilon}/{3H}$.
The quantity, $\Upsilon$ is the dissipation coefficient taken to be a power law in temperature $T$ and the inflaton field, $\Upsilon \propto {T^c}  \phi^{-m}$,  with $c-m=1$. The power indices $c$ and $m$ determine how the inflaton field couples to the radiation fields. 

In the WI paradigm, the slow-roll approximation becomes $3H(1+Q)\dot{\phi} \approx -V_{,\phi}$. The set of slow-roll conditions then include  $Q+1$ terms: 
\begin{eqnarray}
    - \frac{d\,lnH}{dN} = \frac{\epsilon_\phi}{1+Q} < 1~~,\\
    -\frac{d\,ln\,V_{,\phi}}{d\,N} =  \frac{{\eta_\phi}}{1+Q} <1~~,
\end{eqnarray}
where $\epsilon_\phi$ and $\eta_\phi$ are the usual cold inflationary slow-roll parameters, while $ d N\equiv d \ln a$  with $N$  the number of e-folds as the inflaton field evolves, and  $a$ is the scale factor.
Now, imposing the slow-roll approximation, the background equations become:
\begin{eqnarray}
3H(1+Q)\dot{\phi} \approx- V_{,\phi}\, \\
4\rho_R=4C_{R}T^4 \approx 3Q\dot{\phi}^2\, \\
3H^2 \approx \frac{V}{m_P^2} ~~.
\end{eqnarray}
where $C_R= {\pi^2g_*}/{30} $ where $g_*$ is the effective number of relativistic degrees of freedom ($= 106.75$ for standard model particles).

 For simplicity, we adopt cubic dissipation $ \Upsilon= C T^3$ ( $c=3$, $m=0$) with $C$ a constant with  dimension $m_p^{-2}$.
 Although the desired increment in the power spectrum is a generic feature in warm inflation, the amount of enhancement required to associate it with PBHs production seems best suited to cubic dissipation, at least for the case of natural inflation. Of course, this is just one of the possible dissipation form. For example, there can be an explicit field-dependent dissipation form (implicitly the dissipation is always dependent on the field value).  In this case one might imagine that the thermal corrections and  associated effects could  spoil the shift symmetry of the potential. However, this has been addressed in \cite{lwi}, where it was shown that following the prescription of the little Higgs mechanism, one can build up the idea of warm inflation, whereby the inflaton is a pNGB. In this case, the inflaton potential is  protected from the thermal correction due to a cancellation of the correction.

The background equations, along with the definition of Q, can be combined to find a relation between $Q$ and $\phi$ ~\cite{Bastero-Gil:2019rsp}:
\begin{equation}
\label{PhiQrelation}
  \left( 4* 3^{2/c} \right) \left( \frac{ C_R}{C^{4/c}~ m_P^{~2+4/c}} \right ) ~ Q^{4/c-1} (1+Q)^2 =  \left ( \frac{V_{,\phi}^2}{V^{1+2/c}} \right )    
\end{equation}
The evolution of $\phi$, $Q$, and  $T/H$ w.r.t $N$ can be can be obtained using the slow-roll equations and finally for the cubic case one gets:
%%%%%%
\begin{equation}
    \frac{d({T}/{H}) }{dN} =\left( \frac{T}{H} \right) \frac{  \frac{(1-Q)}{4 Q}\frac{dQ}{dN} +\frac{3 \epsilon_{\phi} -\eta_{\phi} }{2}}{1+Q}
\end{equation}
with %%%%%%%%%
\begin{equation}
   \frac{dQ}{dN}=\frac{Q}{1+7Q} \left( 10 \epsilon_{\phi} -6\eta_{\phi} \right)
\end{equation}
so that%%%%%%%%%%%%
\begin{equation}
\frac{d(T/H)}{dN}= \frac{(T/H)}{1+7Q} ~ \left(\frac{8 Q+4}{1+Q} ~\epsilon_{\phi} -2 \eta_{\phi} \right) ~~.
\end{equation}
The power spectrum of curvature perturbations for a non-thermal inflaton distribution is given by (see~\cite{Bartrum:2013fia} and refs. therein):
\begin{equation}
    P_R=\left(\frac{H}{2\pi\dot{\phi}}\right)^2 \left( 1 +\frac{T}{H}F(Q) \right) {G(Q)} ~~, \label{eqPrw1}
\end{equation}
 with
\begin{equation}
    F(Q)
    \approx \frac{2\pi\sqrt{3}Q}{\sqrt{3+4 \pi Q}} ~~\label{eqfQ},
\end{equation}
and
\begin{equation}
    {G(Q)
    =  1 + 4.981 ~ Q^{1.946} + 0.127 ~ Q^{4.330} ~~, }
\label{GQ}
\end{equation}
 where the function $G(Q)$ in Eq.~(\ref{GQ}) accounts for the growth of the inflaton fluctuations due to its coupling with radiation.  This form is a fit to the numerical solution to the full set of perturbation equations found in WI \cite{Benetti17}.
The tensor power spectrum $P_T = 2H^2/(\pi^2 m_p^2)$ is unchanged from that of CI.%%
\begin{center}
\begin{table}
\begin{tabular}{c c c c c c} 
 \hline
 ~~~~Color~~~~ & ~~~~N~~~~ & ~~$\Lambda(m_p^4)$~~ & ~~$C_\phi$~~ & $~~~n_s~~~~$ & ~~~~~~$r$~~~~~ \\ [0.65ex] 
 \hline
 {\color{red} Red} & 55 & $1.00\times 10^{-11}$ & 50 &0.964& $4.0 \times 10^{-3}$ \\ 

 {\color{blue} Blue} & 63 & $7.87\times 10^{-12}$ & 60 &0.966& $4.5 \times 10^{-4}$\\

{\color{green} Green} & 44 & $1.77\times 10^{-12}$ & 40 &0.966& $1.0 \times 10^{-3}$ \\ [.65ex] 
 \hline
\end{tabular}
\caption{(color online) The inflationary observable for different sets of model parameter values. The different color codes are maintained in the plots.}
\label{tab1}
\end{table}
\end{center}
 The enhancement of the power spectrum in comparison to CI can be seen from Eqs.~(\ref{eqPrw1}),  (\ref{eqfQ}) and (\ref{GQ}). The function $F(Q)$ is strictly monotonically  increasing for any $Q (>0)$. Hence, the power spectrum also increases (we note that the term $\frac{T}{H}$  also gives a small  enhancement \cite{Arya:2017zlb}). The major contribution for $Q>1$ comes from $G(Q)$ for cubic dissipation. 
\\

Having formulated the dynamics of WI, we can analyze the power spectrum in the case of WNI. This was first studied in \cite{wnio} and then in \cite{wni1, wni2} The natural inflation potential is given as:
\begin{equation}
    V(\phi)= \Lambda \left( 1+ \cos{\left(\frac{\phi}{f}\right)} \right)~.
\end{equation}

Here, the ``axion" $\phi$ plays the role of an inflaton which is the Goldstone Boson (GB) of the broken symmetry. The factor $f$ is the breaking scale at which the GB  obtains it's potential. In this case, the primordial power spectrum of curvature fluctuations takes the form:
\begin{equation}
\begin{split}
P_{R}= &
\frac{
   \left[ C_{\phi}  \left(~1+ \cos{({\phi}/{f})}~\right) \right]^{4/3} 
   \left({\Lambda }/{m_{P}^4}\right)^{2/3}
   }{48*3^{2/3} ~ \pi ^2 ~ C_{r} ~Q^{1/3}}    
   \\& 
   \left(1+ \frac{3^{2/3}~ Q^{1/3}~ F(Q)}{
    \left[C_{\phi} (~1+ \cos{({\phi}/{f})}~) \right]^{1/3}
   \left({\Lambda }/{m_{P}^4}\right)^{1/6}}
   \right)
   G(Q)~~,
\end{split}
\end{equation}
where for convenience we have taken $C= {C_\phi}/{\Lambda^{1/2}}$ with $C_\phi$ a dimensionless model parameter. 

\begin{figure}[t]
    \includegraphics[width=.45\textwidth, height=.45\textwidth]{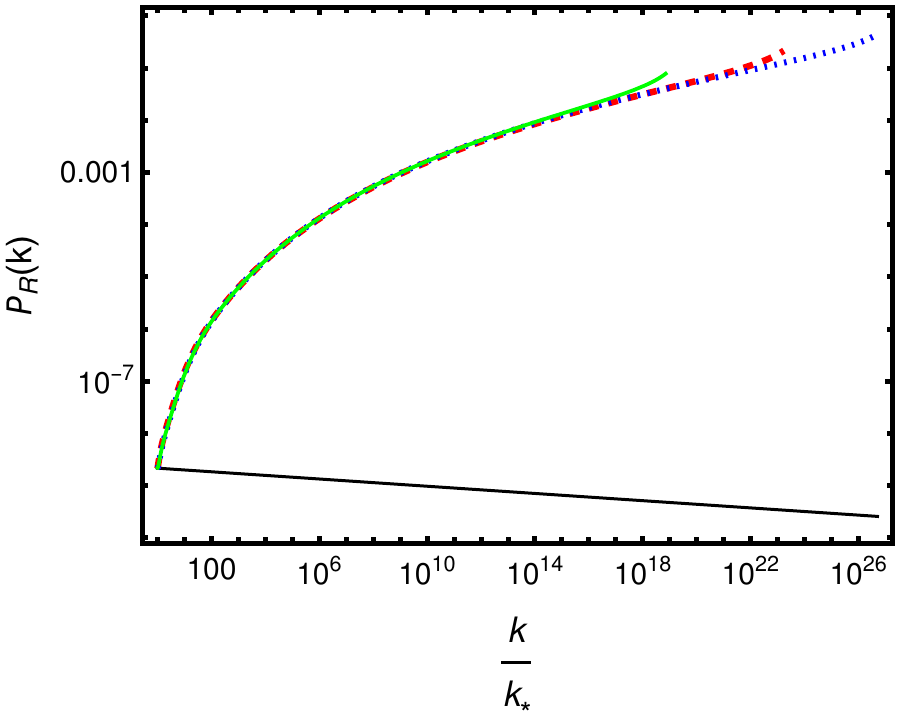}
    \caption{(color online) The curvature power spectrum plotted against the scale ${k}/{k_{\rm *}}$ for the set of parameters with the same color code as in Table~\ref{tab1}, i.e. with the solid green line for $N=44$, the red dashed line for $N=55$, and the blue dotted line is for $N=63$. Here $k_{\rm *}(= 0.02~{\rm Mpc}^{-1})$ is the pivot scale. The enhancement in the spectrum for large k gives the condition for PBH formation. For comparison,  the solid black line at the bottom is the power spectrum for the CI model.}
    \label{fig:P_R}
\end{figure}

The primordial curvature power spectrum for three representative sets of parameters for this model, are shown in Figure \ref{fig:P_R} as a function of wave number scale $k$.   The dependence of $P_R$ on the field value enters through the cosine function.  This affects the enhancement of the power spectrum at relatively smaller $k$ (larger scales) with respect to the previous two analysis \cite{arya, marpbh}. It is also to be noted that, since Q is a function of $H$, there is an additional effect in the power spectrum. 

\subsection{CMB Constraints:}
The three cases studied here
%{\color{red}\sout{,}}  
are summarized in Table \ref{tab1} for different parameter choices.  Setting $f= 0.8~m_p$ for these models leads to CMB consistent results, and {color{red}{it}} ensures that the breaking scale is sub-Planckian. 

The Planck inflation analysis \cite{pl2} has ruled out many inflation models by the constraint that the scalar index for the curvature power spectrum $P_k \propto k^{n_s -1}$, is  $n_s = 0.9649 \pm 0.0042$  at the $68\%$ CL, and more importantly, that the ratio of  the tensor to scalar power spectra has an upper limit of $r \equiv P_T/P_R < 0.056$ at the $68\%$ CL.  Indeed, this result has excluded the natural inflation paradigm in CI if one wants to have $f< m_p$.  However, the three representative parameter sets for the model listed in Table \ref{tab1} easily satisfy the combined constraints on $n_s$ and $r$.  
\\
\\
\section{$Dark~Matter$ PBHs in WNI:}
  PBHs can be formed when large density fluctuations enter the horizon during the radiation-dominated epoch. These density perturbations are quantified as co-moving curvature perturbations. This means that a sufficient enhancement of the primordial power spectrum can provide the  seeds necessary to form  black holes. 
  Large fluctuations are produced during WI at the scale, $k_{\rm PBH}\gg k_*$. The mass of the PBHs depends upon the time of formation and can be parameterized as some fraction $\gamma$ of the horizon mass \cite {Sasaki:2018dmp} \\
  \begin{equation}
      M_{\rm PBH}= \gamma M_H= \gamma \frac{ 4\pi M_{p}^2}{H} \,  , 
  \end{equation}
  where $\gamma$ is  typically taken to be $0.2$ for PBH formation in the radiation dominated era~\cite{Inomata:2017okj}.
  
  Considering the fact that the energy density during the radiation-dominated epoch is  $\rho_r=\frac{\pi^2}{30}g_* T^4$, the PBH mass can written as \cite{Mishra:2019pzq}: \\
  \begin{equation}
      \frac{M_{\rm PBH}}{M_\odot}= 1.13 \times 10^{15}\left (\frac{\gamma}{0.2}\right )\left (\frac{g_*}{106.75}\right )^{-1/6}\left(\frac{k_{\rm PBH}}{k_*}\right )^{-2}\, .
      \label{PBHmasseq1}
  \end{equation}
 
The last factor in the above equation clearly shows that as in most models of WI, the scale associated with   large fluctuations  is largest for small $k_{PBH}$ near the end of inflation leading to  smaller mass PBHs as was the case with the previous two studies. Such small PBHs would form during the last $2-5$ e-folds of inflation. However, a key result in the WNI model is that the enhancement of the fluctuations can be associated with at least $20-25$ e-foldings before the end of  inflation.  This leads to  the successful production of PBHs in the allowed mass range. Indeed,  the dynamics of WI demands an enhancement in the power spectrum.  Until the shift symmetry plays its role, the production of PBHs in the desired mass range would have been a genuine problem. The novelty described here is that WNI naturally resolves this probem.
 
 The present relic abundance of PBHs depends upon  its mass fraction, i.e.~the ratio of the PBH density to the total density at formation. 
 \begin{equation}
     \beta= \left. \frac{ \rho_{\rm PBH}}{\rho_{\rm total}}\right\vert_{\rm formation}\, .
 \end{equation}
 \begin{figure}[htb!]
     \centering
     \includegraphics[height=0.47\textwidth, width=0.53\textwidth]{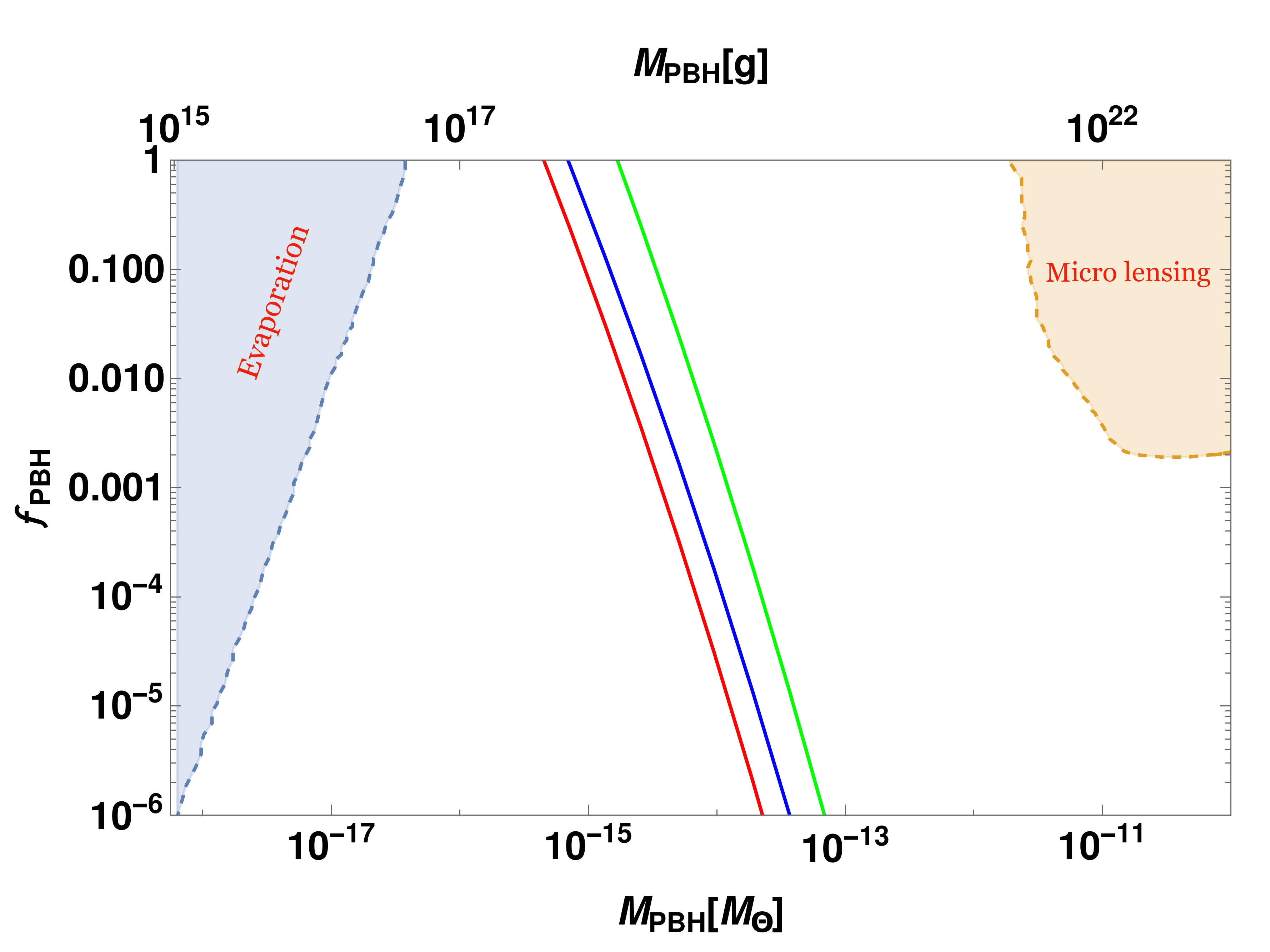}
     \caption{(color online) The fractional abundance of PBH dark matter over the total dark matter as a function of PBH mass (in solar masses and in grams) is shown for the model parameters with the same color code as in Table~\ref{tab1}. All three lines fall into the region of the golden window where a 100\% PBH dark matter contribution is allowed. These plots are also consistent with the bound on $\beta$~\cite{betabound1}}
     \label{abudancef}
 \end{figure}
   PBH formation is considered  to take place during the radiation-dominated epoch. If we assume  that PBHs red shift as non-relativistic matter, then we have the relation\\
  \begin{equation}
      \beta(M_{\rm PBH})= 5.95 \times 10^{-9}\left (\frac{g_*}{106.75}\right )^{\frac{1}{4}}\left(\frac{M_{\rm PBH}}{M_\odot}\right )^{\frac{1}{2}}f_{\rm PBH}\, .
  \end{equation}
 where  $f_{\rm PBH}\equiv {\Omega_{0 \rm PBH 0}}/{\Omega_{0 \rm DM }}$ (also $\gamma=0.2$ is adopted). In Press-Schechter theory \cite{Press:1973iz}, for a Gaussian distribution of the primordial fluctuations,
 the mass function can be written  can as
 \begin{equation}
 \beta(M)={\rm Erc}\left(\frac{\delta_c}{\sqrt{2}\sigma(M)}\right)~~,
 \label{betaeq1}
  \end{equation}
 where ${\rm Erc}$ is the complementary error function, $\sigma ^2 (M)$ is the variance of the fluctuations for the mass scale $M$, and $\delta_c$ is the threshold for the fluctuation.  The variance $\sigma$ directly relates the power spectrum  to the mass fraction. Within  this approximation we can write \cite{Bhattacharya:2021wnk}
 \begin{equation}
     \sigma(M)\simeq \frac{2(1 +w)}{3 + 5 w}\sqrt{P_R (k)}~~.
 \label{sigeq}
 \end{equation}
The initial mass fraction or present abundance  is  sensitive to the  threshold $\delta_c$ (which is taken to be $0.414$ for radiation \cite{Harada:2013epa}) and the curvature power spectrum through Eqs.~(\ref{betaeq1}) and (\ref{sigeq}). Thus, the enhanced power spectrum provided by warm inflation greatly enhances the abundance of PBH dark matter for a particular mass. The mass scale in turn  depends on the scale at which the power spectrum is significant (see Eq.~\ref{PBHmasseq1}). A higher mass PBH with significant contribution to the dark matter density, demands a lower value of $k$, or in other words,  early enhancement of the primordial power spectrum after horizon exit.  This is achieved successfully in our case as shown in  Fig.~\ref{abudancef}.  This figure shows the fraction $f_{PBH}$ of the present dark matter in PBHs vs.~the PBH mass formed in the three parameter sets described in Table \ref{tab1}.  Also shown for comparison are the excluded regions due to black-hole evaporation and observed microlensing events.  All three models fall within the allowed mass range and could account for up to 100\% of the presently observed  dark matter.
%%%%%%%%%%%%%%%%%%%%%%%%%%%%%%%%%%%%%%%%%%%%%%%%%%%%%%%%%%%%%%%%%%%%%%%%%%%%

\section{Conclusion:}
In this letter, we have shown the novel result that in the warm inflationary paradigm if one considers  shift-symmetry-protected natural inflation, the viable fluctuation for producing large-mass PBHs is possible and can account for the whole dark matter energy density by PBHs with a mass range allowed by the black-hole lifetime and observation. In particular, one striking coincidence in this analysis is that  the parameter space allowed by the CMB constraints on the inflationary primordial power spectra is only consistent with the  parameter space required to produce PBHs in the so called golden window whereby the PBHs could account for  the total  observed dark matter density. It would be valuable to justify such an apparent coincidence which we leave to a subsequent work. 

Such an enhancement in the scalar power spectrum suggests a need for a  deeper investigation of the secondary gravitational wave enhancement which is natural from the aspects of perturbation theory. At  second order there is an induced enhancement in the tensor modes. This has been studied first in \cite{wgw1} followed by \cite{marpbh,Domenech:2021ztg,wgw2,Ahmed:2021ucx}. 

Finally, in the case of CI,  an  enhancement in the power spectrum  is difficult to justify in  most cases as one needs to introduce a highly fine tuned inflection point by hand  \cite{cipbh1, cipbh2}.  In the previous two studies it was established that in WI, this enhancement is natural but the absence of large mass PBHs precludes the possibility for such  PBHs to be a viable dark matter candidate. However, we again emphasize in this letter that we have demonstrated the plausible possibility  to produce PBHs in the golden window of mass range, in the well motivated natural inflation scenario utilising the implicit symmetry associated with the breaking and creation of such a potential in the WI paradigm.
\\ 
 \\
%%%%%%
\section*{Acknowledgments:}---
Authors would like to thank M. Sami, Yogesh for valuable suggestions. Work of MRG is supported by DST, Government of India under the Grant Agreement number IF18-PH-228 and partially by Science and SERB, DST, Government of India under the Grant Agreement number CRG/2020/004347.  Work at the University of Notre Dame supported by the U.S. DOE under nuclear theory grant  DE-FG02-95-ER40934. NJ is supported by the NPDF,SERB, DST, Government of India, file No. PDF/2021/004114. NJ is thankful to CCSP, SGT University for hospitality and accommodation where the final draft of this letter is done.

%\bibliographystyle{utphys}
%\bibliography{wipbh}

%%%%%%%%%%
%%%%%%%%%%%%

%%%%%%%%%%%%%%%%%%%%%%%%%%%%%%%%%%%%%%%%%%%%%%%%%%%%%%%%%%%%%%%%%%%%%%%%%%%%%%%

\clearpage
\appendix
%\onecolumngrid
%\section*{Supplemental Material}

%%%%%%%%%%%%%%%%%%%%%%%%%%%%%%%%%%%%%%%%%%%%%%%%%%%%%%%%%%%%%%%%%%%%%%%%%%%%%%%
\end{document}